# Environment-Dependent Components Identification of Behind-the-Meter Resources via Inverse Optimization

Chengming Lyu, Zhenfei Tan, *Member, IEEE*, Xiaoyuan Xu, *Senior Member, IEEE*, Chen Fu,
Zheng Yan, *Senior Member, IEEE*, and Mohammad Shahidehpour, *Life Fellow, IEEE*

*Abstract*—With the increasing penetration of behind-the-meter (BTM) resources, it is vital to monitor the components of these resources and deduce their response behavior to external environment. Owing to data privacy, however, the appliance-wise measurement is invisible to the power system operator, which hinders the accurate modeling of load identification. To this end, this paper proposes a hybrid physics-inspired and data-driven framework for decomposing BTM components based on external measurement of total load and environmental factors. The total load is decomposed into different environment-dependent components, namely storage-like component, PV generation component, thermostatically-controlled load component, and periodic component. The overall load identification adopts a double-layer iterative solution framework. A data-driven inverse optimization algorithm is developed to identify parameters of the energy storage-like component. The physics-inspired model is proposed to identify the capacity and response of the rest components. The modeling accuracy and robustness of the proposed method are validated by numerical tests. The application significance of the proposed BTM identification method is also validated in electricity market clearing for reducing system operation costs.

## Nomenclature

*Abbreviations*

| | |
|---|---|
| BTM | Behind-the-meter |
| ESL | Energy storage-like |
| PV | Photovoltaic |
| TCL | Thermostatically-controlled load |
| PL | Periodic load |
| TL | Total load |
| SO | System operator |

*Parameters*

| | |
|---|---|
| $x_i$ | Environmental factors of the $i$th component |
| $\boldsymbol{P}^{tl}$ | TL power profile |
| $T$ | Time horizon |
| $D$ | Daily time interval number |
| $\pi$ | Electricity price |
| $\gamma$ | Solar irradiance |
| $\tau$ | Outdoor temperature |
| $\boldsymbol{P}^{pv}_{dc}$ | Daily cumulated solar irradiance value |
| $\boldsymbol{P}^{tcl}_{dc}$ | Daily cumulated temperature-load function value |
| $\boldsymbol{P}^{tl}_{dc}$ | Daily cumulated power of the TL |

*Variables*

| | |
|---|---|
| $\theta_i$ | Parameters of the $i$th surrogate model |
| $\boldsymbol{P}^{esl}$ | Power of the ESL component |
| $P^c_i$ | Charging power of the $i$th ESL device |
| $P^d_i$ | Discharging power of the $i$th ESL device |
| $\boldsymbol{P}^{pv}$ | Power of the PV component |
| $\lambda^{pv}$ | Capacity coefficient of the PV component |
| $\boldsymbol{P}^{tcl}$ | Power of the TCL component |
| $\lambda^{tcl}$ | Capacity coefficient of the TCL component |
| $\boldsymbol{P}^{pl}$ | Power of the PL component |
| $\lambda^{pv}_{dc}$ | Initial capacity coefficient of the PV component |
| $\lambda^{tcl}_{dc}$ | Initial capacity coefficient of the TCL component |
| $\sigma^{pl}_{dc}$ | Daily cumulated power of the PL component |
| $\boldsymbol{\theta}$ | Parameter of the virtual battery surrogate model |
| $N$ | Number of virtual batteries |
| $\Omega^{vb}_N(\boldsymbol{\theta})$ | Power feasible region of the virtual battery model |
| $\bar{P}^{vb}_n$ | Power output limit of the $n$th virtual battery |
| $\bar{E}^{vb}_n$ | Upper energy bound of the $n$th virtual battery |
| $\underline{E}^{vb}_n$ | Lower energy bound of the $n$th virtual battery |
| $\boldsymbol{P}^{vb}_n$ | Power trajectory of the $n$th virtual battery |

## I. Introduction

UNDER the global trend of energy decarbonization, the increasing penetration of renewable energy, e.g., photovoltaic (PV) generation and wind power, poses a great challenge to the operation of power system. The conventional operation mode that generators follow the fluctuation of load and renewable generation is no longer applicable. In addition, behind-the-meter (BTM) resources, i.e., user-owned energy storage systems, electric vehicles, and flexible loads, are becoming increasingly important to the balance of the renewable-dominated modern power system. Due to data privacy and management boundaries, power system operators (SO) can merely measure the total net load demand of each user, but cannot directly monitor or dispatch the BTM resources [1]. The BTM resources include a broad spectrum of devices with diversified operation patterns. The load or generation of these devices may be impacted by multiple external factors, including electricity price, air temperature, and solar irradiation. How to identify different BTM components of an active user and model their responses to external factors is a fundamental problem of utilizing the demand-side flexibility.

Existing studies regarding load component identification fall into two categories, i.e., microscopic methods and macroscopic methods. The microscopic method aims at monitoring what electric appliances are owned by a user as well as their load profiles. Both intrusive load monitoring (ILM) and non-intrusive load monitoring (NILM) can be classified into microscopic load modeling. As for ILM, the Pecan Street is a representative example [2], which measures circuit-level electricity use from hundreds of residential houses and offers datasets specific to electric vehicle charging, rooftop solar generation, energy storage, and energy use down to individual household appliances [3]. The database shows great sufficiency and reliability and can be used as a benchmark in comparison with newly proposed load monitoring methods [4]. However, ILM usually requires a large number of sensors and considerable installation costs when accessing new devices, and the collected information is certainly considered private [5]. As a contrast, NILM deduces the activities of appliances from the total power of the consumer [6]. Since only the total power consumption is measured and device-level consumption is estimated through decomposition algorithms, NILM offers advantages of low cost, easy implementation, and scalable commercialization [7].

For integrated energy systems or household scenarios, traditional NILM decomposes the load precisely to certain devices by extracting distinctive load features and matching them with the typical load characteristics database [8]. Information of active power [9], voltage-current trajectory [10], and harmonic content [11] are commonly utilized. Moreover, machine learning algorithms, e.g., artificial neural network [12] and hidden Markov model [13], have continuously deepened the research on NILM. Furthermore, latest research in deep learning and big data analytics has enabled data-driven approaches leveraging large-scale datasets [14]. Accordingly, notable approaches based on deep learning, e.g., the convolutional neural network [15], long short time memory network [16], and the denoising auto encoder [17], have been applied in the literature. Though ILM and NILM can monitor what electric appliances are working at different times, these microscopic load decomposition methods cannot model how the demand-side resources respond to external variables such as spot market price. In addition, these methods cannot measure the potential flexibility of demand-side resources, which is critical for the operation of power system.

In contrast, macroscopic load modeling studies the load as an aggregated entity and derives demand response characteristics from historical data. These studies analyze load profiles at various scales, including individual users, distribution networks, and regional systems. Among these studies, the most common application is the load forecasting. Typical methods include time series analysis, machine learning methods, and deep learning algorithms [18]. Considering the reaction of electricity consumers to the external environment, reference [19] proposed a hybrid forecasting framework that incorporates dynamic electricity price effects to model consumer response. Using a data-driven probabilistic net load forecasting method, reference [20] models the relationship between the net load and solar radiation. Based on the ZIP static load model, reference [21] disaggregates power consumption of low-voltage substations into thirteen representative categories according to their characteristics of the aggregate active and reactive power. Correspondingly, reference [22] estimates the shares of different load categories and overall controllable and non-controllable load within the total forecasted load, with a foreseen application in various demand response programs. On this basis, reference [23] estimates flexible loads with different elasticity characteristics from the substation scale and decomposes loads into interruptible and deferrable components.

Through the substation-wise load composition, reference [24] estimates the percentage of industrial loads, commercial loads, and residential loads. The macroscopic load model, however, is indistinct for the SO to accurately dispatch the power system with the presence of active load. For instance, the SO dispatches the system and clears the electricity market based on the forecasted load. But the actual load may deviate from the forecasted one according to the price signal, which may result in systematic deviation in load forecasting. Therefore, it is essential to model the response behavior of active load, rather than only forecasting. Nevertheless, load models based on general machine learning, such as the neural network and support vector regression, are nonlinear and cannot be directly embedded into the optimal dispatch model of power system.

Hence, an ideal load behavior model must satisfy three critical requirements, i.e., 1) interpretable: the net load can be decomposed into several components related to external variables; 2) deducible: the load response can be predicted with the information of external variables; and 3) embeddable: the load model can be directly embedded into the power system optimal dispatch problem without posing additional computation burden. The aforementioned microscopic models do not possess interpretability and deducibility, while the macroscopic ones are not embeddable. To this end, a new load modeling task, namely the environment-dependent component identification (EDCI), is studied in this paper. The EDCI aims at decomposing the total load (TL) profile of an active user into key components that respond to different environmental factors, e.g., electricity price, solar irradiance, and temperature. To realize the EDCI with the aforementioned three requirements, a hybrid physics-inspired and data-driven framework is developed to identify load components from historical data of TL and environmental factors.

In this framework, the data-driven inverse optimization method is proposed to identify the capacity and parameters of energy storage-like (ESL) resources. The physics-inspired model is proposed to identify the capacity of the PV component and the thermostatically-controlled load (TCL) component. The contribution of this paper is twofold,

1) A novel hybrid physics-inspired and data-driven framework is developed for identifying environment-dependent BTM components. By constructing proper surrogate models and determining model parameters, the proportion and response behavior of each component can be precisely estimated. Compared to the microscopic and macroscopic modeling approaches, the proposed physics-inspired and data-driven method ensures component-level interpretability, response deducibility, and practical embeddability.

2) To solve the EDCI problem that is cast into a bilevel programming problem, a double-layer iterative solution framework combining physics-inspired load models and data-driven inverse optimization method is proposed. In particular, the response behavior of PV and TCL is identified based on the physics-inspired model. The virtual battery fleet model is used as a surrogate model whose response is trained to closely resemble the BTM ESL component through the data-driven inverse optimization method. The resulted load component model is linear and will not bring additional computation burden in power system optimal dispatch problem.

In the rest of this paper, the problem formulation and the solution algorithm of the EDCI will be introduced in Section II and Section III, respectively. The numerical simulations and result analysis will be provided in Section IV. Section V will conclude this work.

## II. Problem Formulation

### A. Framework

As shown in Fig. 1, different types of BTM resources have diversified responses to the external environment factors. For example, a user-side energy storage system will optimally determine its charge and discharge activities based on the electricity price. The power output of a PV generation is highly dependent on the solar irradiance. The air-conditioning load demand is influenced by the environmental temperature. The SO can acquire the total power profile data through smart meters at coupling points of users. However, the SO does not have access to detailed capacity and parameter information about BTMs. To assist the SO monitor the BTM components, the EDCI problem is conceptually formulated as follows,

$$\min_{f_i, \theta_i} \left\| \sum_i f_i(x_i; \theta_i) - \boldsymbol{P}^{tl} \right\|_2 \tag{1}$$

The goal of the EDCI is to identify a set of surrogate models whose aggregated power response to environmental factors can mimic the measured power trajectory of active users. In equation (1), the surrogate model $f_i(x_i; \theta_i)$ characterizes the response behavior of the $i$th BTM component of the user, where $x_i$ denotes environmental factors that impact the power trajectory of component $i$. $\theta_i$ represents the parameters of the surrogate model. $\boldsymbol{P}^{tl}$ is the TL measured by smart meters.

The EDCI problem involves two tasks, i.e., 1) designing a proper surrogate model for each BTM component, and 2) determining parameters of surrogate models to reduce the modeling error. The surrogate model should be interpretable and can be encoded in the subsequent optimal dispatch problem of power system. In this paper, the form of each surrogate model is designed through the physical understanding of BTM resources. The parameters of each surrogate model are identified via an iterative solution framework.

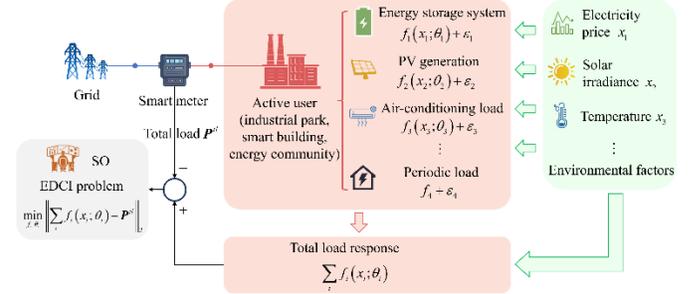

Fig. 1. Framework of the EDCI problem.

The EDCI problem does not focus on disaggregating the load into specific devices. Instead, it emphasizes on identifying response behaviors of different BTM resources. The identified component can be used to predict the load response to external factors and thus benefit the power supply-demand balance.

### B. Physical Modeling of BTM Components

Four typical BTM load components are considered, i.e., the ESL resources that respond to the electricity price, the PV generation that responds to the solar irradiance, the TCL component that responds to the outdoor temperature, and the periodic load (PL) with a fixed daily periodic pattern. Surrogate models of these components are established based on their physical models.

1) ESL component

Consider a fleet of user-side ESL resources indexed by $i \in \mathcal{L}$ that optimally schedule their power trajectories according to the day-ahead electricity price. Given the day-ahead electricity price series $\boldsymbol{\pi} \in \mathbb{R}^T$ over the time horizon $T$, the aggregated power consumption $\boldsymbol{P}^{esl} \in \mathbb{R}^T$ of ESL devices is determined by the following optimization problem,

$$\boldsymbol{P}^{esl} = \sum_{i \in \mathcal{L}} \left( \boldsymbol{P}_i^c - \boldsymbol{P}_i^d \right) \tag{2}$$

$$\left( \boldsymbol{P}_i^c, \boldsymbol{P}_i^d \right) = \arg\min \boldsymbol{\pi}^\top \left( \boldsymbol{P}_i^c - \boldsymbol{P}_i^d \right) \tag{3}$$

$$\text{s.t. } F_i \left( \boldsymbol{P}_i^c, \boldsymbol{P}_i^d \right) \le 0, \tag{4}$$

$$\boldsymbol{P}_i^c \ge 0, \boldsymbol{P}_i^d \ge 0. \tag{5}$$

$\boldsymbol{P}_i^c$ and $\boldsymbol{P}_i^d$ denote the charging and discharging power vectors of the $i$th ESL device, respectively. Equation (3) describes the objective function of the ESL power optimization problem. Equation (4) enforces the operation constraints of ESL devices, including power limits, energy limits, and ramping limits, and etc.

2) PV component

As validated in [25], the output of PV generation is linearly proportional to the solar irradiance level. Accordingly, in the surrogate model of the PV component, the total power output of the BTM PV $\boldsymbol{P}^{pv} \in \mathbb{R}^T$ is determined by

$$\boldsymbol{P}^{pv} = \sum_{j \in \mathcal{M}} \boldsymbol{P}^{pv}_j = \lambda^{pv} \cdot \boldsymbol{\gamma}, \quad (6)$$

where $j \in \mathcal{M}$ is the index of PV. $\boldsymbol{\gamma} \in \mathbb{R}^T$ denotes the time-varying series of solar irradiance and $\lambda^{pv}$ is a coefficient proportional to the installed capacity of the PV.

3) TCL component

The equivalent thermal parameter model is used to characterize the thermal control of each electrical heating/cooling load [26], i.e.,

$$\frac{d\tau_i}{dt} = \frac{1}{RC}|\tau_o - \tau_i| - \frac{Q^t}{C}, \quad (7)$$

where $\tau_i$ and $\tau_o$ represent the indoor and outdoor temperature, respectively. $R$ and $C$ denote the equivalent thermal resistance and capacitance. $Q^t$ indicates the thermal output produced by the TCL. Under the assumption that the TCL can keep a constant indoor temperature, i.e., $d\tau_i/dt = 0$, the thermal output $Q^t$ produced by the TCL is as follows,

$$Q^t = \frac{1}{R}|\tau_o - \tau_i|. \quad (8)$$

According to [27], the TCL electric power $P^{tcl}$ has a linear function relationship with its compressor operating frequency $f$. The thermal output $Q^t$ is a quadratic function of $f$.

$$P^{tcl} = a \cdot f + b, \quad (9)$$

$$Q^t = c \cdot f^2 + d \cdot f + e, \quad (10)$$

where $a, b, c, d, e$ are constant coefficients and $c$ is negative. Based on (9) and (10), the relationship between electric power and temperature is as follows,

$$P^{tcl} = a \cdot \frac{d \pm \sqrt{d^2 - 4ce + \frac{4c|\tau_o - \tau_i|}{R}}}{-2c} + b. \quad (11)$$

Considering the value of $d$ and $c$, only the negative sign can yield a physically meaningful solution in (11). Under constant indoor temperature regulation, the TCL electric power can be simplified as a function of the outdoor temperature, denoted as $P^{tcl} = g(\tau_o)$.

In the surrogate model of the TCL component, the aggregated power of a group of TCLs can be estimated by scaling up $g(\tau_o)$ with a capacity factor, i.e.,

$$\boldsymbol{P}^{tcl} = \sum_{k \in \mathcal{N}} \boldsymbol{P}^{tcl}_k = \lambda^{tcl} \cdot g(\boldsymbol{\tau}), \quad (12)$$

where $\boldsymbol{P}^{tcl} \in \mathbb{R}^T$ denotes the power trajectory of TCL and $\boldsymbol{\tau} \in \mathbb{R}^T$ is a vector composed of the outdoor temperature $\tau_o$.

4) PL component

In addition to the aforementioned environment-dependent load components, the PL component is independent with external factors and has a fixed daily-periodic pattern. In this way, the TL behavior can be estimated by the summation of the four load components.

*C. Mathematical Formulation of EDCI Problem*

Given historical measurements of total power consumption $\boldsymbol{P}^{tl}$ and environmental factors, the target of the EDCI is to find a set of surrogate models to decompose the TL into different components. The EDCI problem is formulated as follows,

$$\min_{\theta, \lambda^{pv}, \lambda^{tcl}, \boldsymbol{P}^{esl}, \boldsymbol{P}^{pv}, \boldsymbol{P}^{tcl}, \boldsymbol{P}^{pl}} \ell = \left\| \boldsymbol{P}^{esl} + \boldsymbol{P}^{pv} + \boldsymbol{P}^{tcl} + \boldsymbol{P}^{pl} - \boldsymbol{P}^{tl} \right\|_2 \quad (13)$$

$$\text{s.t.} \quad \boldsymbol{P}^{esl} = \arg\min\left\{ \boldsymbol{\pi}^\top \cdot \boldsymbol{P}^{esl} : \boldsymbol{P}^{esl} \in \Omega(\boldsymbol{\theta}) \right\}, \quad (14)$$

$$\boldsymbol{P}^{pv} = \lambda^{pv} \cdot \boldsymbol{\gamma}, \quad (15)$$

$$\boldsymbol{P}^{tcl} = \lambda^{tcl} \cdot g(\boldsymbol{\tau}), \quad (16)$$

$$\boldsymbol{P}^{pl}(n+D) = \boldsymbol{P}^{pl}(n), \forall n \leq T - D, n \in \mathbb{N}^+. \quad (17)$$

In (14), $\Omega(\boldsymbol{\theta})$ denotes the feasible region of the surrogate model parameterized by vector $\boldsymbol{\theta}$. $\boldsymbol{P}^{esl}$ is the optimal response under the electricity price $\boldsymbol{\pi}$. $\lambda^{pv}$ and $\lambda^{tcl}$ are the identified coefficients representing the installed capacity of PV and TCL, respectively. The periodic load component is specified in (17), where $D$ denotes the daily time interval number. The objective in (13) is to minimize the estimation error between the response of the surrogate load model and the measured load profile.

The EDCI is realized following two steps, i.e., 1) designing a proper surrogate model $\Omega(\boldsymbol{\theta})$ of the ESL, 2) solving the optimization problem to determine parameters $\boldsymbol{\theta}, \lambda^{pv}, \lambda^{tcl}$ and $\boldsymbol{P}^{pl}$. To track the evolvement of user-side system structure, e.g., the installation of new devices, the EDCI problem can be solved and updated on a weekly or multi-day basis.

### III. METHODOLOGY

*A. Solution Framework*

The EDCI formulation in (13)-(17) constitutes a bilevel programming problem. It is commonly converted into a single-layer problem based on the KKT condition or value function. However, this direct solving method faces computational challenges due to non-linear constraints and the combinatorial complexity introduced by the conversion. To this end, an iterative solution framework is developed for the efficient solution of the EDCI.

As shown in Fig. 2, the iterative solution framework involves four steps. Firstly, based on the TL data and the external environment data, the PV and TCL components are estimated as the initial value. Secondly, the surrogate model parameters of the ESL component are deduced by solving a reverse optimization problem. Thirdly, coefficients of PV and TCL are updated through an equality-constrained quadratic programming problem. Finally, the PL component is obtained by subtracting the other components from the TL. These four steps are repeated until the calculation result converges or the maximum iteration number is reached.

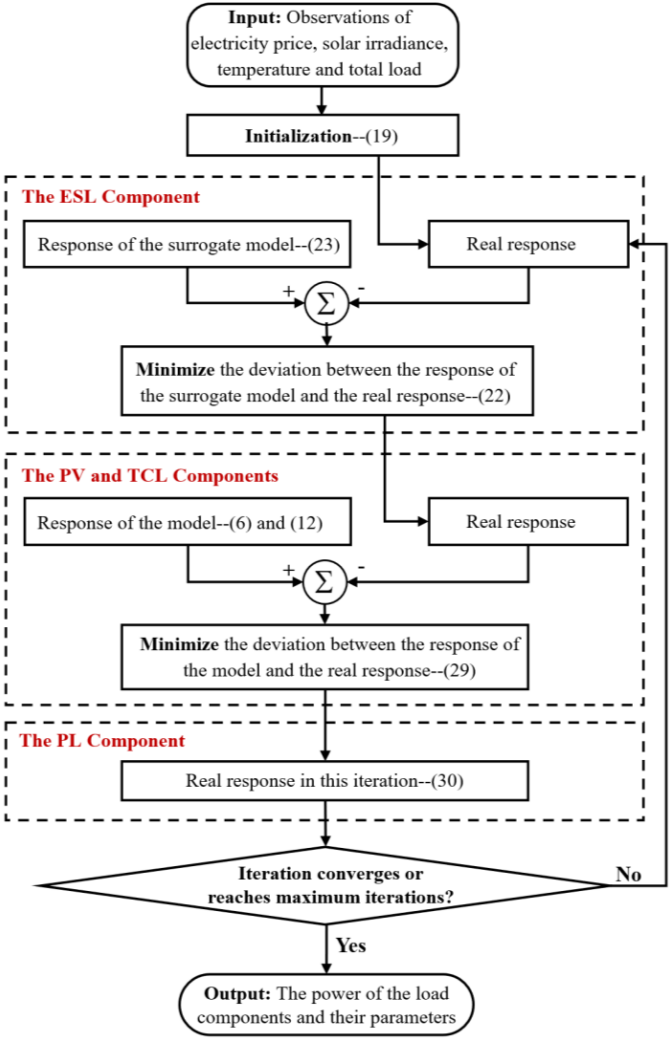

Fig. 2. Iterative solution framework.

*B. Initialization*

An initial guess of the PV, TCL, and PL components can be obtained through the initialization based on two physical constraints: 1) the cumulated power of the ESL component in one day is 0, and 2) the cumulated power of the PL component in one day is a constant (denoted as $\sigma_{dc}^{pl}$). By performing a $D$-point summation, the relationship between the daily cumulant of the TL and its components is as follows:

$$\sum_{n=1}^{D} P^{tl}(n) = \sum_{n=1}^{D} P^{esl}(n) + \sum_{n=1}^{D} P^{pv}(n) + \sum_{n=1}^{D} P^{tcl}(n) + \sum_{n=1}^{D} P^{pl}(n) \\ = 0 + \sum_{n=1}^{D} P^{pv}(n) + \sum_{n=1}^{D} P^{tcl}(n) + \sigma_{dc}^{pl}. \quad (18)$$

Conducting the same summation on the solar irradiance data and the temperature-load function $g(\tau_o)$, the initial iteration value of PV and TCL can be obtained by solving the following problem,

$$\min_{\sigma_{dc}^{pl}, \lambda_{dc}^{pv}, \lambda_{dc}^{tcl}} \ell_i = \left\| \lambda_{dc}^{pv} \cdot \boldsymbol{P}_{dc}^{pv} + \lambda_{dc}^{tcl} \cdot \boldsymbol{P}_{dc}^{tcl} + \sigma_{dc}^{pl} - \boldsymbol{P}_{dc}^{tl} \right\|_2 \quad (19)$$

s.t. $P_{dc}^{pv}(m) = \sum_{n=(m-1)D+1}^{mD} \gamma(n),$

$P_{dc}^{tcl}(m) = \sum_{n=(m-1)D+1}^{mD} g(\tau(n)),$ (20)

$P_{dc}^{pl}(m) = \sum_{n=(m-1)D+1}^{mD} P^{pl}(n), \forall m \in \mathbb{N}^+, mD \leq T.$

Problem (19) is an equality-constrained quadratic programming problem, which can be effectively solved through the gradient descent method. Dividing the PL daily cumulant by the daily time interval number, the initial guess of the PL component in different time intervals is a constant, i.e., $\sigma_{dc}^{pl}/D$.

*C. Surrogate Model for ESL Component*

To capture the response behavior of the ESL component, a set of virtual batteries is used as the surrogate model, i.e.,

$$\Omega_N^{vb}(\boldsymbol{\theta}) = \left\{ \boldsymbol{P}^{esl} \left| \begin{array}{c} \boldsymbol{P}^{esl} = \sum_n \boldsymbol{P}_n^{vb}, \\ -\overline{\boldsymbol{P}}_n^{vb} \leq \boldsymbol{P}_n^{vb} \leq \overline{\boldsymbol{P}}_n^{vb}, \\ \underline{\boldsymbol{E}}_n^{vb} \leq \Lambda \boldsymbol{P}_n^{vb} \leq \overline{\boldsymbol{E}}_n^{vb}, n = 1:N \end{array} \right. \right\}. \quad (21)$$

In this model, $\boldsymbol{\theta} = (\ldots, \overline{P}_n^{vb}, \overline{E}_n^{vb}, \underline{E}_n^{vb}, \ldots)^\top \in \mathbb{R}^{3N}$ is the parameter to be determined, which includes the power output limits $\overline{P}_n^{vb}$, upper energy capacity bounds $\overline{E}_n^{vb}$, and lower energy capacity bounds $\underline{E}_n^{vb}$. $N$ denotes the number of virtual batteries and is a tunable hyperparameter for model fidelity. $\boldsymbol{P}_n^{vb} \in \mathbb{R}^T$ denotes the power trajectory of the $n$th virtual battery. $\Lambda$ is a unit lower triangular $T \times T$ matrix and $\Lambda \boldsymbol{P}_n^{vb}$ yields the vector of the accumulated energy by each time interval, which represents the state-of-charge of the virtual battery. This linear model can be seamlessly integrated into optimization problems related to system dispatch and aggregation strategies.

Given the real response of the ESL component and electricity price, the determination of the surrogate model parameters $\boldsymbol{\theta}$ constitutes an inverse optimization problem:

$$\min_{\boldsymbol{\theta}, \boldsymbol{P}^{esl}} \ell_g = \left\| \boldsymbol{P}^{esl} - \left( \boldsymbol{P}^{tl} - \lambda^{pv} \cdot \boldsymbol{\gamma} - \lambda^{tcl} \cdot g(\boldsymbol{\tau}) - \boldsymbol{P}^{pl} \right) \right\|_2 \quad (22)$$

s.t. $\boldsymbol{P}^{esl} = \arg\min \left\{ \boldsymbol{\pi}^\top \cdot \boldsymbol{P}^{esl} : \boldsymbol{P}^{esl} \in \Omega_N^{vb}(\boldsymbol{\theta}) \right\}. \quad (23)$

The objective of the inverse optimization problem is to minimize the estimation error between the power-electricity price response of the surrogate model and the real response of the ESL component.

*D. Newton-Based Inverse Optimization Algorithm*

By leveraging the mathematical structure of (22), a Newton-based inverse optimization algorithm is proposed to solve the bilevel programming problem. By reformulating the coefficients in (21) into matrices as $A$, $B$, and $C$, $\Omega_N^{vb}(\boldsymbol{\theta})$ can be substituted into (14) and rewritten as

$$\boldsymbol{P}^{esl}(\boldsymbol{\theta}) = \arg\min \left\{ \boldsymbol{\pi}^\top \cdot \boldsymbol{P}^{esl} : A\boldsymbol{P}^{esl} + B\boldsymbol{P}^{vb} + C\boldsymbol{\theta} \leq 0 \right\}. \quad (24)$$

Problem (24) is a multi-parametric linear programming [28]. The feasible region of $\theta$ can be divided into several non-overlapping critical regions and the optimal solution $P^{esl}(\theta)$ is a linear and affine function regarding $\theta$ [29]. By calculating $P^{pv}$, $P^{tcl}$ and $P^{pl}$ based on the initial iteration value and substituting $P^{esl}(\theta)$ into (22), equation (22) becomes a non-constrained programming problem. This can be solved through the Newton method by updating $\theta$ iteratively. It can be deduced from the first-order condition that

$$\frac{\partial \ell_g}{\partial \theta} = \frac{\partial \ell_g}{\partial P^{esl}} \frac{\partial P^{esl}}{\partial \theta}$$
$$= 2\left[P^{esl}(\theta) + \lambda^{pv} \cdot \gamma + \lambda^{tcl} \cdot g(\tau) + P^{pl} - P^{tl}\right]^\top \frac{\partial P^{esl}}{\partial \theta} \quad (25)$$
$$= 0.$$

At the $k$th iteration, assume $\theta = \theta^{(k)}$. The optimal solution of the ESL component power in problem (22) can be solved as $P^{esl}(\theta^{(k)})$. In the critical region containing $\theta^{(k)}$, the binding constraints at the optimum stay invariant, which can be denoted as

$$\overline{A}^{(k)} P^{esl} + \overline{B}^{(k)} P^{vb} + \overline{C}^{(k)} \theta = 0. \quad (26)$$

For degenerate problems, this equation is deterministic and $P^{esl}$ can be solved as

$$\begin{bmatrix} P^{esl} \\ P^{vb} \end{bmatrix} = -\begin{bmatrix} \overline{A}^{(k)} & \overline{B}^{(k)} \end{bmatrix}^{-1} \overline{C}^{(k)} \theta \triangleq \begin{bmatrix} F^{(k)} \theta \\ G^{(k)} \theta \end{bmatrix}. \quad (27)$$

Substituting (27) into (25), the updating formula of $\theta$ can be derived as

$$2\left[F^{(k)}\theta^{(k+1)} + \lambda^{pv} \cdot \gamma + \lambda^{tcl} \cdot g(\tau) + P^{pl} - P^{tl}\right]^\top F^{(k)} = 0 \Rightarrow$$
$$\theta^{(k+1)} = \left[F^{(k)\top} F^{(k)}\right]^{-1} F^{(k)\top} \left(P^{tl} - \lambda^{pv} \cdot \gamma - \lambda^{tcl} \cdot g(\tau) - P^{pl}\right). \quad (28)$$

The parameter identification algorithm of the surrogate model is shown as Algorithm 1. To make this algorithm locally convergent, $F^{(k)\top} F^{(k)}$ should be positive definite, which can be generally satisfied in practice if $T \gg 3N$. The computational cost of this algorithm is low, for it merely needs to solve a series of linear programming problems in (24). As for the initial value of $\theta^0$, a proper guess close to the global optimum can be generated based on a grid fitting method [30].

*E. Double-Layer Iteration Algorithm*

The complete EDCI algorithm employs a double-layer iterative structure, i.e., the ESL component is identified in the inner loop, and the other load components are identified in the outer loop. In the $l$th outer iteration loop, the power response of the ESL component $P^{esl(l)} = P^{esl}(\theta^*)$ is calculated in the inner loop through Algorithm 1. Then, the PV and TCL components are identified through the following quadratic optimization problem (29),

$$\min_{\lambda^{pv(l)}, \lambda^{tcl(l)}} \ell_o = \left\| \lambda^{pv(l)} \cdot \gamma + \lambda^{tcl(l)} \cdot g(\tau) + P^{esl(l)} + P^{pl(l-1)} - P^{tl} \right\|_2. \quad (29)$$

After obtaining the coefficients of PV and TCL, the PL can be updated as

$$P^{pl(l)} = P^{tl} - P^{esl(l)} - \lambda^{pv(l)} \cdot \gamma - \lambda^{tcl(l)} \cdot g(\tau). \quad (30)$$

The iteration converges when the successive estimations of the PL component satisfy:

$$\frac{\left\| P^{pl(l)} - P^{pl(l-1)} \right\|_2}{\left\| P^{pl(l)} \right\|_2} \leq \upsilon, \quad (31)$$

where $\|\cdot\|$ denotes the $l_2$-norm and $\upsilon$ denotes the pre-set accuracy value. The entire algorithm of EDCI is stated in Algorithm 2.

---

**Algorithm 1** Surrogate Model Parameter Identification

**Input:** Observations of $P^{tl}$ and estimations of $P^{pv}$, $P^{tcl}$ and $P^{pl}$.
**Output:** The parameters and response of the virtual battery-based surrogate model.
**Initialization:** $k = 0$ and $\theta^{(k)} = \theta^0$.
**while** $k < K$ **do**
　solve (24) with $\theta^{(k)}$ and calculate coefficients $\overline{A}^{(k)}$, $\overline{B}^{(k)}$ and $\overline{C}^{(k)}$ of binding constraints.
　update $\theta^{(k+1)}$ via (28) and calculate $\ell_g^{(k)}$ and $\ell_g^{(k+1)}$ via (22).
　**if** $\left|\ell_g^{(k+1)} - \ell_g^{(k)}\right| \leq \varepsilon$, **break.**
　set $k = k+1$.
**end while**
**return** $\theta^{(k+1)}$ and $P^{esl}(\theta^{(k+1)})$.

---

**Algorithm 2** Iterative Solution for EDCI

**Input:** Observations of $P^{tl}$, $\pi$, $\gamma$, $\tau$ and function $g(\tau_o)$.
**Output:** Different environment-dependent load components and their parameters.
**Initialization:** Calculate the initial iteration value of the PV, TCL, and PL based on (19)-(20). Set $l = 0$, $\lambda^{pv(0)} = \lambda^{pv}_{dc}$, $\lambda^{tc(0)} = \lambda^{tc}_{dc}$, $P^{pl(0)}(n) = \sigma^{pl}_{dc}/D, \forall n \leq T, n \in \mathbb{N}^+$.
**while** $l < L$ **do**
　Execute **Algorithm 1** to solve the response of the ESL component $P^{esl(l)}$.
　Solve the PV coefficient $\lambda^{pv(l)}$ and the TCL coefficient $\lambda^{tcl(l)}$ via **(29)**.
　Calculate the PL power $P^{pl(l)}$ via **(30)**.
　**if** (31), **break.**
　set $l = l+1$.
**end while**
**return** $\theta^{(k+1)}, \lambda^{pv(l)}, \lambda^{tcl(l)}, P^{esl}(\theta^{(k+1)}), P^{pv(l)}, P^{tcl(l)}, P^{pl(l)}$.

## IV. NUMERICAL TEST

The effectiveness of the proposed method is validated through numerical tests by comparing with other mainstream machine-learning methods. The electricity price data $\pi$ and temperature data $\tau$ are collected from the ISO New England [31], where the hourly day-ahead locational price and dry-bulb temperature at load zone NEMA in 2022 are used. To keep geographical consistency with the NEMA load zone, the solar irradiance data $\gamma$ in Boston is used and acquired from the NASA data access viewer [32]. These external data sets are sampled on an hourly basis and thus $D$ is 24.

To generate data set for simulation, the TL data is made up of four components. The ESL component $\boldsymbol{P}^{esl}$ is generated by solving (2)-(5) on a system with 50 heterogeneous responsive loads with operation constraints formulated in [33]. Each responsive load has a power capacity of 4 MW and an energy capacity randomly sampled from 8 to 24 MW. The PV component $\boldsymbol{P}^{pv}$ and TCL component $\boldsymbol{P}^{tcl}$ are calculated through (6) and (12) respectively, with carefully tuned coefficients $\lambda^{pv}$ and $\lambda^{tcl}$ to maintain a reasonable proportion. The PL component $\boldsymbol{P}^{pl}$ is generated by summing the real power consumption of several industrial companies. The summation retains random errors to validate the method's generalization ability. The experimental setup employs a sliding-window approach, i.e., 6 consecutive days for training followed by 3 subsequent days for testing. Simulations are conducted in MATLAB R2022b with GUROBI V10.0 solver on a laptop with an Intel Core i7-12700H processor.

The normalized root mean square error (NRMSE) metric is used to evaluate the model's performance.

$$NRMSE = \frac{\sqrt{\frac{1}{N}\sum_{i=1}^{N}(\tilde{y}_i - y_i)^2}}{\max|y_i|} \times 100\%,$$

where $y_i$ and $\tilde{y}_i$ denote actual and estimated values, respectively. The normalization is conducted by dividing the RMSE metric by the maximum absolute value to reflect the relative range of error fluctuations.

In Section IV-A, the error reduction trajectories of different load components during the iteration process are visualized. In Section IV-B, the comparative performance analysis of the proposed method is presented against benchmark machine learning methods. In Section IV-C, the effect of reducing dispatching cost through EDCI is validated by embedding the result into the unit commitment optimization.

### A. Illustration of Solution Process

In each iteration of the proposed algorithm, the four load components are sequentially identified. Fig. 3 illustrates the convergence process of NRMSEs across iterations. Due to the existence of estimation error, the initial iteration values of the PV and TCL components exhibit relatively high errors. However, as the iteration progresses, the NRMSEs of different load components consistently decrease and eventually stabilize, demonstrating the convergence of the iteration process.

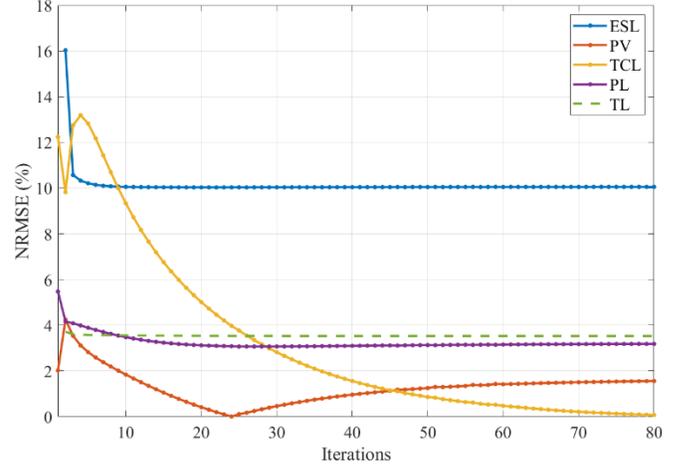

Fig. 3. Illustration of NRMSEs decreasing with iterations.

### B. Load Component Identification Results

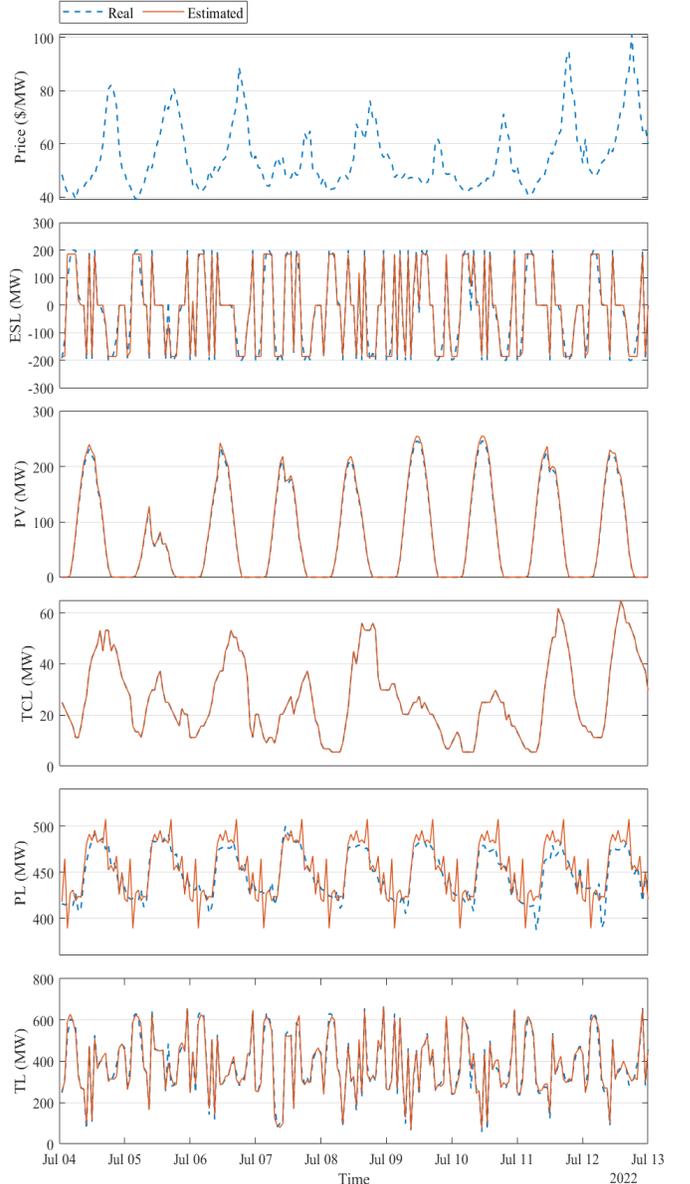

Fig. 4. Comparison of modeling performance.

As illustrated in Fig. 4, the modeling performance of the proposed method is demonstrated by comparing estimated and actual values of the load components. The evaluation employs a training set of July 4-9 and a test set of July 10-13. Notably, the first row in Fig. 4 reveals a significant difference in the electricity price curve between the training set and the test set. Therefore, this period is selected for exhibition to examine the applicability of the proposed method under adverse conditions. Rolling tests are conducted throughout July using a sliding window approach: each 9-day segment is divided into 6 training days followed by 3 test days. The average NRMSEs of load components and TL are listed in Table I. The results of both the training set and the test set show small errors, indicating that the proposed method exhibits a satisfactory fitting performance.

TABLE I
NRMSEs OF LOAD COMPONENTS

| Load Component | Training NRMSE | Test NRMSE |
|---|---|---|
| TL | 3.05% | 4.03% |
| PL | 4.68% | 4.63% |
| ESL | 9.13% | 14.98% |
| PV | 5.25% | 5.26% |
| TCL | 4.14% | 4.04% |

### C. Accuracy Comparison

To evaluate the effectiveness of the proposed method, mainstream machine learning methods, including the multilayer perceptron (MLP) and support vector machine (SVM), are used as benchmarks for comparison. Similar to the rolling window test in Section B, simulations are conducted for 31 periods with the initial date spanning July 1-31. The training set includes 6 consecutive days, and the test set includes the subsequent 3 days. The electricity price, temperature, and solar irradiance data are employed as inputs. The TL data is served as the output of the machine learning methods.

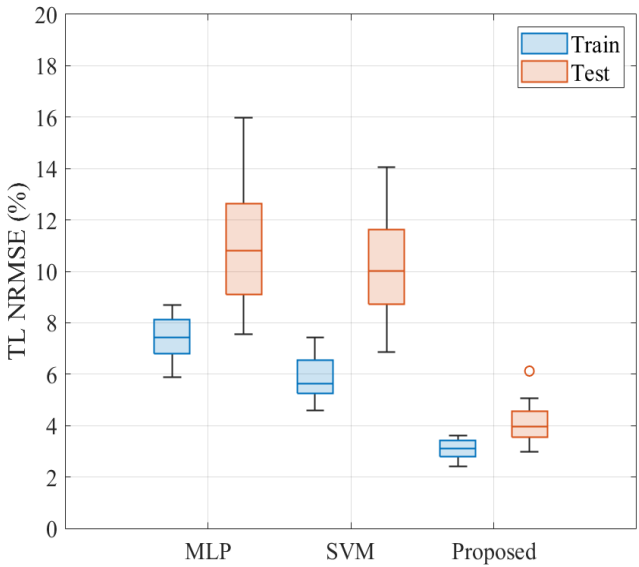

Fig. 5. Comparison of the proposed method with machine learning methods.

Hyperparameters of these models are finely tuned to improve accuracy on both training and test sets. Since the load components are not explicitly involved in machine learning methods, only the NRMSE of the TL is presented in Fig. 5. The proposed method demonstrates superior performance and achieves NRMSE values below 3.7% (training set) and 5.1% (test set). In contrast, the MLP and SVM methods show significantly higher errors: the training NRMSEs range from 4.6% to 8.7%, and the test NRMSEs range from 7.6% to 16.0%. Notably, the NRMSE distribution of the proposed method on both sets is also more concentrated, indicating more robust performance compared to the machine learning methods.

### D. Application Result in Optimal Dispatch

The power system dispatch and electricity market clearing rely on load forecasting accuracy. However, the real-time load may deviate from the forecasted one after the market clearing price is published. It is necessary to adjust the output of the units or even change the on-off state, resulting in considerable adjustment costs. Based on the proposed EDCI method, the installed capacities of the ESL, PV, and TCL components can be estimated. The proposed method and the SVM method are used to predict the load, respectively, where the training set is the TL, electricity price, irradiance, and temperature data for 6 consecutive days. The costs of the two methods are illustrated in Fig. 6 and Table II. As can be seen, the day-ahead generation costs of the two methods are similar, while the real-time adjustment cost of the proposed method is remarkably reduced by 58.4%. The superior forecast accuracy of our method minimizes the need for generation output adjustments, particularly eliminating costly commitment status changes. Overall, these improvements yield a 17.6% reduction in total system dispatch costs, demonstrating the significant economic benefits of the proposed method in power system operations.

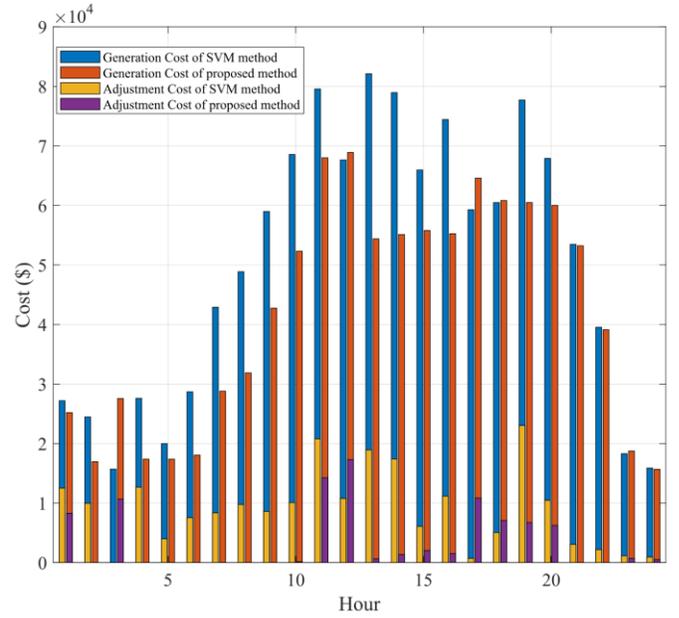

Fig. 6. Performance of the proposed method in electricity market clearing

TABLE II
COST OF THE TWO METHODS ($)

| Method | Day-ahead Generation Cost | Real-time Adjustment Cost | Unit Start-stop Cost | Total Cost |
|---|---|---|---|---|
| SVM method | 988648 | 215450 | 20000 | 1224098 |
| Proposed method | 919958 | 88384 | 0 | 1008342 |

## V. Conclusion

In this paper, a hybrid physics-inspired and data-driven load identification method is presented, which decomposes the TL into different environment-dependent components, including the PL, ESL, PV, and TCL components. A set of surrogate models is designed to characterize their response behaviors. An iterative solution framework with an initialization technique and a Newton-based inverse optimization algorithm is developed. Numerical tests validate the modeling performance, and comparison with the MLP and SVM methods shows the accuracy and robustness of the proposed method. Its application in electricity market clearing also reveals its effectiveness in reducing generation and adjustment costs.